\documentclass[10pt,a4paper,twocolumn,superscriptaddress,aps,prd,longbibliography,nofootinbib]{revtex4-2}
\usepackage[utf8]{inputenc} 
\usepackage[T1]{fontenc}    
\usepackage{lmodern}        

\usepackage[modulo]{lineno}
\usepackage{amsmath}
\usepackage{amsfonts}
\usepackage{amssymb}
\usepackage{bbold}
\usepackage{graphicx}
\usepackage[usenames,dvipsnames]{color}
\usepackage{float}
\usepackage{multirow}
\usepackage{soul}
\usepackage{hyperref}
\usepackage{ulem}

\usepackage[nohyperlinks, printonlyused, nolist]{acronym}

\usepackage{verbatim} 

\graphicspath{ {./images/} }

\usepackage{blindtext}

\begin{document}

\author{Cedric Schmitt}
\thanks{These authors have contributed equally}
\affiliation{Physikalisches Institut, Universit\"at W\"urzburg, D-97074 W\"urzburg, Germany}
\affiliation{W\"urzburg-Dresden Cluster of Excellence ctd.qmat, Universit\"at W\"urzburg, D-97074 W\"urzburg, Germany} 

\author{Lukas Gehrig}
\thanks{These authors have contributed equally}
\affiliation{Physikalisches Institut, Universit\"at W\"urzburg, D-97074 W\"urzburg, Germany}
\affiliation{W\"urzburg-Dresden Cluster of Excellence ctd.qmat, Universit\"at W\"urzburg, D-97074 W\"urzburg, Germany}

\author{Jonas Erhardt}
\affiliation{Physikalisches Institut, Universit\"at W\"urzburg, D-97074 W\"urzburg, Germany}
\affiliation{W\"urzburg-Dresden Cluster of Excellence ctd.qmat, Universit\"at W\"urzburg, D-97074 W\"urzburg, Germany}

\author{Kilian Strau\ss}
\affiliation{Physikalisches Institut, Universit\"at W\"urzburg, D-97074 W\"urzburg, Germany}
\affiliation{W\"urzburg-Dresden Cluster of Excellence ctd.qmat, Universit\"at W\"urzburg, D-97074 W\"urzburg, Germany}

\author{Stefan Enzner}
\affiliation{W\"urzburg-Dresden Cluster of Excellence ctd.qmat, Universit\"at W\"urzburg, D-97074 W\"urzburg, Germany}
\affiliation{Institut f\"ur Theoretische Physik und Astrophysik, Universit\"at W\"urzburg, D-97074 W\"urzburg, Germany}

\author{Martin Kamp}
\affiliation{Physikalisches Institut, Universit\"at W\"urzburg, D-97074 W\"urzburg, Germany}
\affiliation{Physikalisches Institut and R\"ontgen Center for Complex Material Systems, D-97074 W\"urzburg, Germany}

\author{Timur Kim}
\affiliation{Diamond Light Source, Harwell Science and Innovation Campus, Didcot, OX11 0DE, United Kingdom}

\author{Giorgio Sangiovanni}
\affiliation{W\"urzburg-Dresden Cluster of Excellence ctd.qmat, Universit\"at W\"urzburg, D-97074 W\"urzburg, Germany}
\affiliation{Institut f\"ur Theoretische Physik und Astrophysik, Universit\"at W\"urzburg, D-97074 W\"urzburg, Germany}

\author{J\"org Sch\"afer}
\affiliation{Physikalisches Institut, Universit\"at W\"urzburg, D-97074 W\"urzburg, Germany}
\affiliation{W\"urzburg-Dresden Cluster of Excellence ctd.qmat, Universit\"at W\"urzburg, D-97074 W\"urzburg, Germany}

\author{Simon Moser}
\affiliation{Physikalisches Institut, Universit\"at W\"urzburg, D-97074 W\"urzburg, Germany}
\affiliation{W\"urzburg-Dresden Cluster of Excellence ctd.qmat, Universit\"at W\"urzburg, D-97074 W\"urzburg, Germany}
\affiliation{Experimentalphysik IV - AG Oberflächen, Ruhr-Universität Bochum, 44801 Bochum, Germany}

\author{Ralph Claessen}
\email{e-mail: claessen@physik.uni-wuerzburg.de}
\affiliation{Physikalisches Institut, Universit\"at W\"urzburg, D-97074 W\"urzburg, Germany}
\affiliation{W\"urzburg-Dresden Cluster of Excellence ctd.qmat, Universit\"at W\"urzburg, D-97074 W\"urzburg, Germany}

\date{\today}


\title{Enhanced Screening in Epitaxial Graphene via Nearly Free-Electron Metal Intercalation}

\maketitle

\noindent \textbf{
Graphene exhibits extraordinarily high carrier mobility, making it a promising platform for next-generation electronics. Scalable growth on SiC, however, suffers from limited dielectric screening at the graphene–substrate interface, degrading electronic performance. In this work, we systematically enhance dielectric screening by intercalating a bilayer of indium at the graphene–SiC interface. Using graphene’s plasmaronic signature observed in angle-resolved photoemission spectroscopy as a proxy for interaction strength, we quantitatively demonstrate strong dielectric screening arising from the interplay of both indium layers. Layer-resolved density functional theory shows that the first indium layer acts as a buffer that absorbs substrate interactions, enabling the second layer to form a nearly free-electron system that efficiently screens the graphene layer above. Experiments with only a single intercalated indium layer reveal reduced screening, confirming the essential role of the second layer. Our results establish 2ML indium intercalation as a powerful route for engineering dielectric environments in graphene.}


\section*{}
\begin{figure*}[t!]
\includegraphics[width=18cm, height=25cm, keepaspectratio]{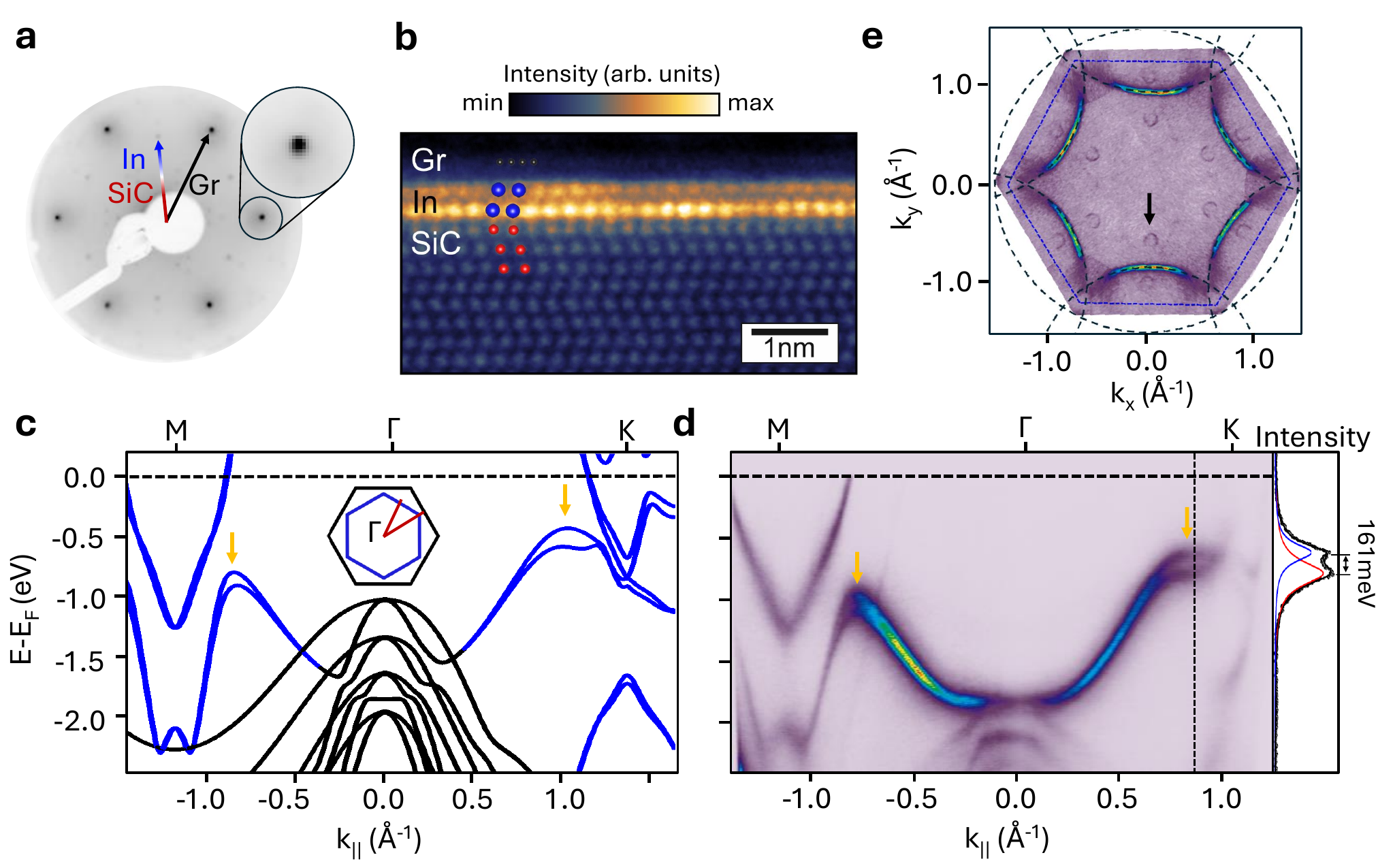}
\caption{\textbf{Intercalated bilayer indium} \textbf{a} LEED image taken at 80$\,$eV showing the diffraction spots of 2ML indium $(1 \, \times \,1)$ (blue arrow) and graphene (black). A zoom-in to the latter demonstrates indium intercalation through the absence of the characteristic ($6\sqrt{3} \, \times \,6\sqrt{3}$)R30$^\circ$ periodicity of the ZLG. \textbf{b} RT STEM image of intercalated indium revealing positions of Si (of SiC, red spheres), indium (blue spheres) and graphene (black spheres). \textbf{c} DFT calculations of pristine bilayer indium band structure along the high-symmetry directions. 
A Rashba-type splitting induced by SOC is clearly visible (orange arrows), with the bands splitting along both $\Gamma K$ and $\Gamma M$ directions. \textbf{d} ARPES band structure recorded at 10$\,$K with $h\nu=40$ eV. The highlighted energy distribution curve (dashed line in the ARPES spectra) is fitted with two Voigt profiles, yielding a separation of 161 meV between the Rashba-split bands. \textbf{e} Bilayer indium Fermi surface built from 6-fold symmetrization of data in \textbf{d} showing the characteristic signature of a nearly free-electron-like In band (dashed black circles). The black arrow marks the graphene replica, which is consistent with electron diffraction from the In/SiC lattice.}
\label{fig:BilayerIndium}
\end{figure*}

Graphene has emerged as a paradigmatic two-dimensional electron system, combining atomic-scale thickness with exceptional electronic properties \cite{doi:10.1126/science.1102896,Kim2009,https://doi.org/10.1002/admt.202000744}. From the earliest stages of its experimental exploration, charge transport in graphene stood out for exhibiting remarkable high carrier mobilities \cite{doi:10.1126/science.1102896,PhysRevLett.100.016602}. This demonstration of ultrafast carrier dynamics established graphene as a prime platform for studying low-dimensional quantum transport and for pursuing high-performance electronic and optoelectronic functionalities.

In realistic device environments, however, the transport properties of graphene are strongly degraded by extrinsic scattering mechanisms of their surrounding environment \cite{Newaz2012,https://doi.org/10.1002/adma.201805656,PhysRevLett.101.146805,PhysRevB.77.115449}. Graphene quality, substrate and interface effects drive the carrier mobility far below its intrinsic potential and thus limit device performance \cite{ma16247668}. Despite these obstacles, high mobilities exceeding $2 \times 10^{5}~\mathrm{cm^{2}V^{-1}s^{-1}}$ have been achieved in suspended graphene, highlighting the intrinsic quality of the material in the absence of a supporting substrate \cite{BOLOTIN2008351,PhysRevLett.101.096802}. Yet, an alternative strategy, which further enhances the mobility, exploits dielectric environment engineering: interfacing graphene with different layers can efficiently screen the long-range Coulomb interactions \cite{Chen2009,PhysRevLett.102.206603,Wei2016,10.1063/1.3077021,Zhang2022}. For instance, graphene encapsulated between hexagonal boron nitride (hBN) layers and placed in close proximity (1\,nm) to a metallic gate exhibits strong electrostatic screening via image-charge effects \cite{Domaretskiy2025}. This proximity screening suppresses long-range potential fluctuations and charge inhomogeneity, resulting in a dramatic enhancement of electronic quality, with transport mobilities exceeding $10^{8}\,\mathrm{cm^2\,V^{-1}\,s^{-1}}$ \cite{Domaretskiy2025}.

Despite these advances, practical applications require wafer-scale material platforms. Large-area graphene suitable for technological integration is presently realized almost exclusively by epitaxial growth on SiC substrates \cite{deHeer_2010,Riedl_2010}. Although this approach provides uniform, scalable graphene, strong substrate interactions substantially limit the attainable mobilities. Even after hydrogen intercalation, which decouples the graphene layer from the substrate, typical mobilities remain on the order of $1.1 \times 10^{4}~\mathrm{cm^{2}V^{-1}s^{-1}}$, far below the intrinsic regime \cite{Pallecchi2014,10.1063/1.3643034,Robinson2011}.

\textcolor{black}{Intercalation of atomic species at the graphene--SiC interface has emerged as a powerful strategy for engineering the structural, electronic, and dielectric properties of epitaxial graphene while enabling the synthesis of atomically thin materials \cite{Ag,Riedl,Gehrig,Pb1,Pb2,Sb,Au_Intercalation,Viro,Bisti,Schmitt,SchmittRaman,Walter}. By inserting foreign atoms between graphene and the SiC substrate, the graphene--substrate interaction can be controlled, enabling wafer-scale growth of high-quality monoelemental two-dimensional layers \cite{Ag}. Depending on the intercalated species, these systems exhibit diverse emergent phenomena, including two-dimensional topological insulating phases \cite{Gehrig,Pb1,Schmitt,SchmittRaman}, semiconductor-to-metal transitions \cite{Au_Intercalation}, and tunable dielectric environments that modify electronic screening and many-body interactions \cite{Walter}.} \textcolor{black}{In this context, indium intercalation is particularly versatile, encompassing indium oxide \cite{InOxide}, indium alloys \cite{InGa}, and elemental indium \cite{Briggs2020,10.1039/d3na00630a,HU2021829,10.1063/5.0223972,KIM2020229,BLIN}, whose thickness can be controlled from a monolayer to a bilayer, a capability that is central to the present study.} 
\textcolor{black}{Here, we focus on graphene screening via intercalation of a bilayer indium to the graphene/SiC interface.}  By adapting to the SiC surface periodicity, the first indium layer saturates dangling bonds and buffers the substrate-induced mirror-symmetry breaking, manifested in a pronounced Rashba-type spin splitting \cite{Bauernfeind,Schmitt}. This allows the second indium layer to develop nearly free-electron-like states at the Fermi surface, thereby achieving exceptionally strong dielectric screening of the graphene sheet. 

To demonstrate and evaluate the screening capability of bilayer indium, we employ ARPES to probe the well-established many-body interactions at the Dirac point of graphene. These include electron--phonon coupling, manifested in a characteristic loss feature near the Fermi level \cite{Aaron2,McChesney}, as well as electron--plasmon interactions that give rise to plasmaron excitations \cite{Lundqvist1967} and the associated plasmaronic band structure \cite{Aaron1,Polini,Hwang,Hwang2,Hwang3}. The latter provides quantitative insight into dielectric screening and quasiparticle dynamics in two-dimensional systems \cite{Walter}.

\begin{figure*}[t!]
\includegraphics[width=18cm, height=25cm, keepaspectratio]{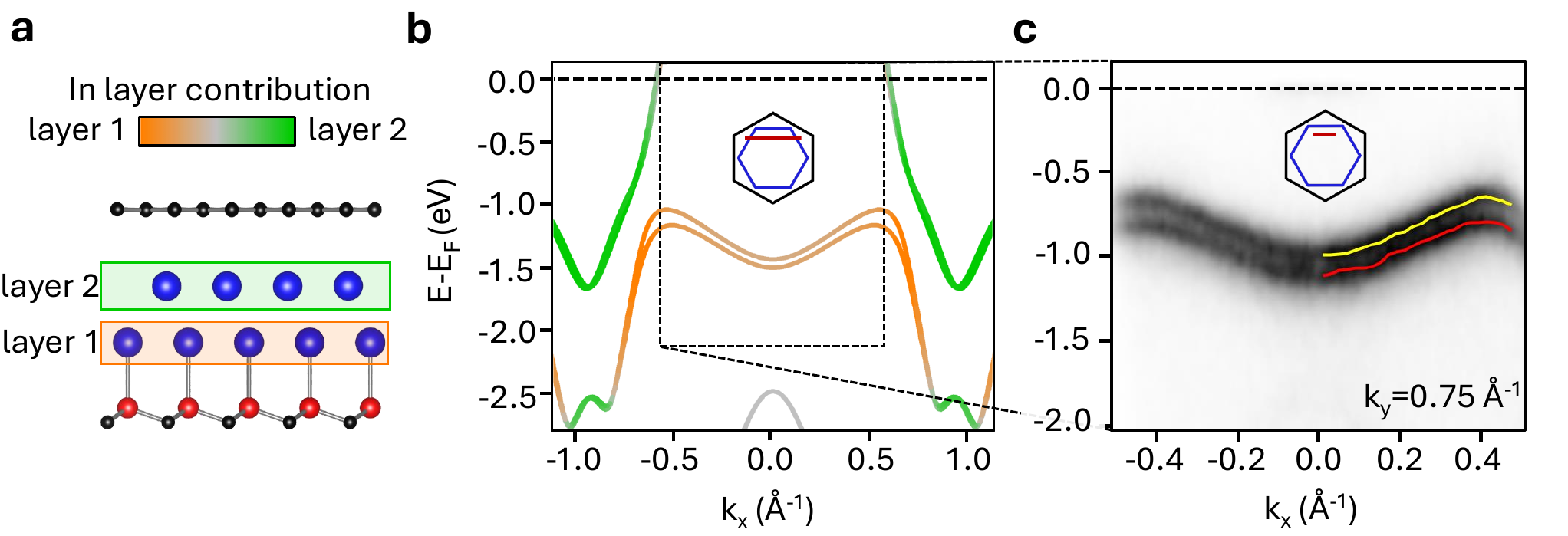}
\caption{\textbf{Layer-resolved indium contribution} \textbf{a} Schematic ball-and-stick model for intercalated bilayer indium. The contribution of the first indium layer is shown in orange, and that of the second layer in green. \textbf{b} DFT band structure of bilayer indium calculated along the Brillouin zone path, shown with the layer-resolved indium contribution in color. \textbf{c} Zoomed-in experimental ARPES band structure recorded at 10$\,$K with $h\nu=40\,$eV together with the corresponding BZ schematic. Voigt-profile fits to the EDCs spectra reveal a large Rashba splitting between the two bands.}
\label{fig:Rashba}
\end{figure*}

The growth process of this heterostructure begins with the formation of zero-layer graphene (ZLG) on SiC(0001), produced by high-temperature Si sublimation, yielding the characteristic $(6\sqrt{3} \times 6\sqrt{3})$R30$^\circ$ surface reconstruction. During indium intercalation, ZLG is displaced as the bonding partner of the SiC substrate, effectively lifted and transformed into a single layer of
quasi freestanding graphene. \textcolor{black}{After intercalation, the $(6\sqrt{3} \times 6\sqrt{3})R30^\circ$ periodicity is strongly suppressed in low energy electron diffraction (LEED) (Fig.~\ref{fig:BilayerIndium}a), leaving predominantly graphene and $(1 \times 1)$ indium diffraction spots (for further details see Supporting Information).}

Representative scanning transmission electron microscopy (STEM) measurements (Fig.~\ref{fig:BilayerIndium}b) reveal two projected indium layers, each containing one In atom per surface Si atom. This configuration implies a $(1 \times 1)$ adsorption geometry of a $2$ ML In film on SiC(0001), consistent with the absence of additional LEED superperiodicities and related studies \cite{Briggs2020}. \textcolor{black}{Corresponding first-principles equilibrium-phase stability calculations reported in Ref.~\cite{Briggs2020} indicate that thicker In films are energetically unfavorable beneath graphene, as additional In atoms are more favorably accommodated within a three-dimensional bulk-like phase rather than incorporated into the confined interfacial film.} High-resolution STEM further resolves the adsorption sites of the two layers: the lower layer occupies the T1 site above the topmost Si atoms, similar to intercalation of a monolayer indium \cite{Schmitt}, while the upper layer resides at the T4 site with respect to the SiC substrate.

\begin{figure*}[t!]
\includegraphics[width=18cm, height=25cm, keepaspectratio]{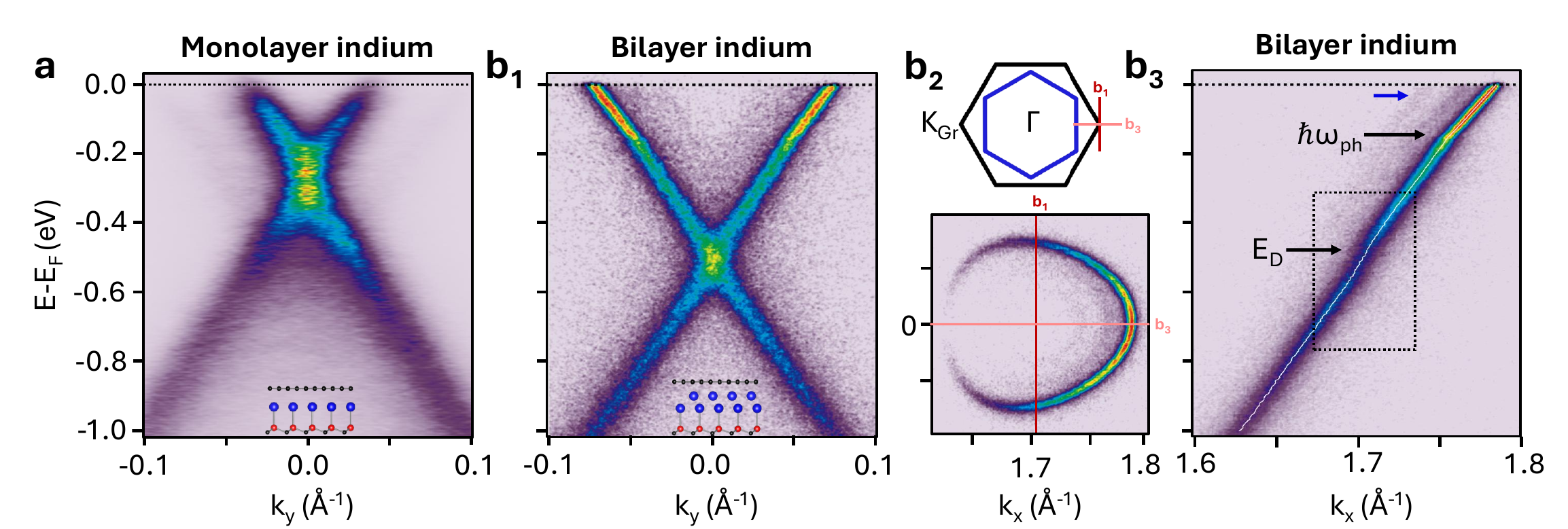}
\caption{\textbf{Plasmaron excitations in graphene.} ARPES spectra recorded at 10$\,$K with $h\nu=46\,$eV near the graphene $K_{Gr}$ point for intercalated \textbf{a} 1ML indium and \textbf{b} 2ML indium. Two orthogonal cuts for graphene on bilayer indium are shown in \textbf{b\textsubscript{1}} and \textbf{b\textsubscript{3}}, with the Fermi surface map displayed in \textbf{b\textsubscript{2}}.
The blue arrow indicates a subtle signature of a differently doped graphene layer, arising from small monolayer indium regions. The white line in Fig. \textbf{b\textsubscript{3}} indicates the band dispersion obtained from the peak positions of the Lorentzian-fitted MDCs.}
\label{fig:Graphene}
\end{figure*}

Turning to the electronic properties of this heterostructure, Fig.~\ref{fig:BilayerIndium}c and d presents density functional theory (DFT) band structure calculated for pristine bilayer indium on SiC next to the corresponding ARPES dispersion, demonstrating excellent agreement between theory and experiment. In addition to bands reminiscent of the quantum spin Hall insulator indenene, \textit{i.e.}, a monolayer of indium on SiC \cite{Bauernfeind,Schmitt,Erhardt2024,Erhardt2025}, an In $sp$ band \cite{Erhardt2022} emerges around the indium $M$ point. These metallic bands exhibit an exceptionally large Fermi velocity of approximately $(1.13 \pm0.18) \times 10^{6}~\mathrm{m\,s^{-1}}$, exceeding that of graphene's $\pi$ states \cite{PhysRevLett.103.186802,Zhang2005} and being comparable to In on Si(111) \cite{PhysRevB.92.041410} and bilayer Ga intercalated at the graphene/SiC interface \cite{Briggs2020}. This nearly free electron like band forms circular Fermi surface pockets that are backfolded at the Brillouin-zone boundary (see Fig.~\ref{fig:BilayerIndium}e). Similar bands are observed for bilayer indium on Si(111) \cite{PhysRevLett.91.246404}, where the first In layer screens the Si substrate and the second layer exhibits nearly free electron behavior \cite{PhysRevLett.117.116102,PhysRevLett.109.166102}. The black arrow in the ARPES spectra highlights the graphene replica, whose position is consistent with electron diffraction originating from the In/SiC lattice. 

Further, our high resolution data reveal a pronounced Rashba-type spin splitting of $161\,$meV at local bandmaxima in $\Gamma$K and $\Gamma$M direction, as indicated by yellow arrows and an EDC in Fig.~\ref{fig:BilayerIndium}d \textcolor{black}{(see Supporting Information for further details)}. This value is comparable with recent measurements of spin-split bands in trilayer gallium \cite{Yi2026}.
This spin-splitting is most pronounced along the energy–momentum cut presented in Fig.~\ref{fig:Rashba}, and indicates a strong susceptibility of these indium bands to the surface potential of the underlying SiC, similar to Shockley surface states on noble metals \cite{PETERSEN200049,PhysRevLett.77.3419, Park2011}.
Corresponding layer-resolved DFT calculations reveal that this strongly Rashba-split band originates from the first indium layer (highlighted in orange in Fig.~\ref{fig:Rashba}a), whereas negligible splitting is observed for the metallic nearly free bands associated with the second indium layer (green in Fig.~\ref{fig:Rashba}b).
This finding demonstrates the role of the first indium layer as a buffer, whose atoms mask the strong surface potential of the substrate, thereby providing an ideal environment for the nearly free electrons in the second layer and thus screening for graphene above.


Having established the properties of intercalated bilayer indium, we now turn to a quantitative ARPES analysis of graphene’s screening. To highlight the relevance of both indium layers in this mechanism, we directly compare graphene on bilayer and monolayer indium intercalation in the following. 
In both cases, ARPES confirms the decoupling of graphene from the SiC substrate upon indium intercalation  (monolayer in Fig.~\ref{fig:Graphene}a and bilayer indium in Fig.~\ref{fig:Graphene}b), as evidenced by the characteristic linear Dirac band crossing at the graphene $K_{Gr}$ point. A constant-energy map of graphene intercalated with bilayer indium at $E_\mathrm{F}$ is shown in Fig.~\ref{fig:Graphene}b\textsubscript{2}, with the two orthogonal momentum cuts corresponding to the dispersions displayed in Fig.~\ref{fig:Graphene}b\textsubscript{1} and Fig.~\ref{fig:Graphene}b\textsubscript{3}. The characteristic horseshoe-shaped intensity distribution of graphene is clearly observed, including the dark corridor arising from the structure factor of the honeycomb lattice \cite{Shirley1995,Daimon1995,Aaron2,Mucha2008,Gierz2011,PhysRevResearch.5.033075}. In the following, we demonstrate how the substrate screening strength can be quantitatively extracted from the plasmaronic many-body interactions of graphene using the bilayer indium system as a benchmark. An analogous analysis is performed for the monolayer case.

We note in passing that our high-resolution spectra reveal the characteristic band renormalization arising from electron–phonon coupling, manifested as a kink in the dispersion at $\hbar \omega_\mathrm{ph}~=~(162~\pm~5)~\mathrm{meV}$ below the Fermi level. The uncertainty is estimated from the energy resolution of the experimental setup, as the error obtained from the min–max analysis is smaller than the instrumental energy resolution. This energy corresponds to the $A_1'$ optical phonon mode ($160~$meV) at the $K_{Gr}$ point, involving in-plane carbon vibrations that efficiently couple to Dirac carriers \cite{PhysRevB.76.205411}. Linear fits above and below the kink yield the characteristic renormalization energy, in agreement with previous observations \cite{Aaron2,McChesney}. The electron-phonon coupling strength can be extracted from the mass-enhancement relation \cite{Grimvall}, which simplifies for linear band dispersions to $\lambda = v_{\mathrm{ph}}/v_{\mathrm{F}} - 1$, where $v_{\mathrm{ph}}$ denotes the band velocity below the phonon kink \cite{BOSTWICK200763}. The resulting electron--phonon coupling strength of $\lambda = 0.2161 \pm 0.0026$ falls well within the range of experimentally reported values for graphene \cite{BOSTWICK200763}.

Returning to the analysis of graphene screening, we identify a characteristic band renormalization around the Dirac point associated with plasmaron excitations—quasiparticles formed by the coupling of a photohole to a plasmon of nearly matched velocity \cite{Aaron1}. The momentum and energy offsets between the primary hole band and the satellite plasmaron band provide a direct measure of the effective dielectric screening in graphene \cite{Aaron1,Walter}. The hole and plasmaron bands are schematically shown in the inset of Fig.~4a. In addition to the crossing of the hole energy bands ($E_1$), a second crossing appears, originating from the plasmaron bands ($E_2$). The separation between $E_1$ and $E_2$ indicates the strength of the coupling between plasmons and charge carriers in graphene.

\begin{figure}[t!]
\includegraphics[width=1\columnwidth]{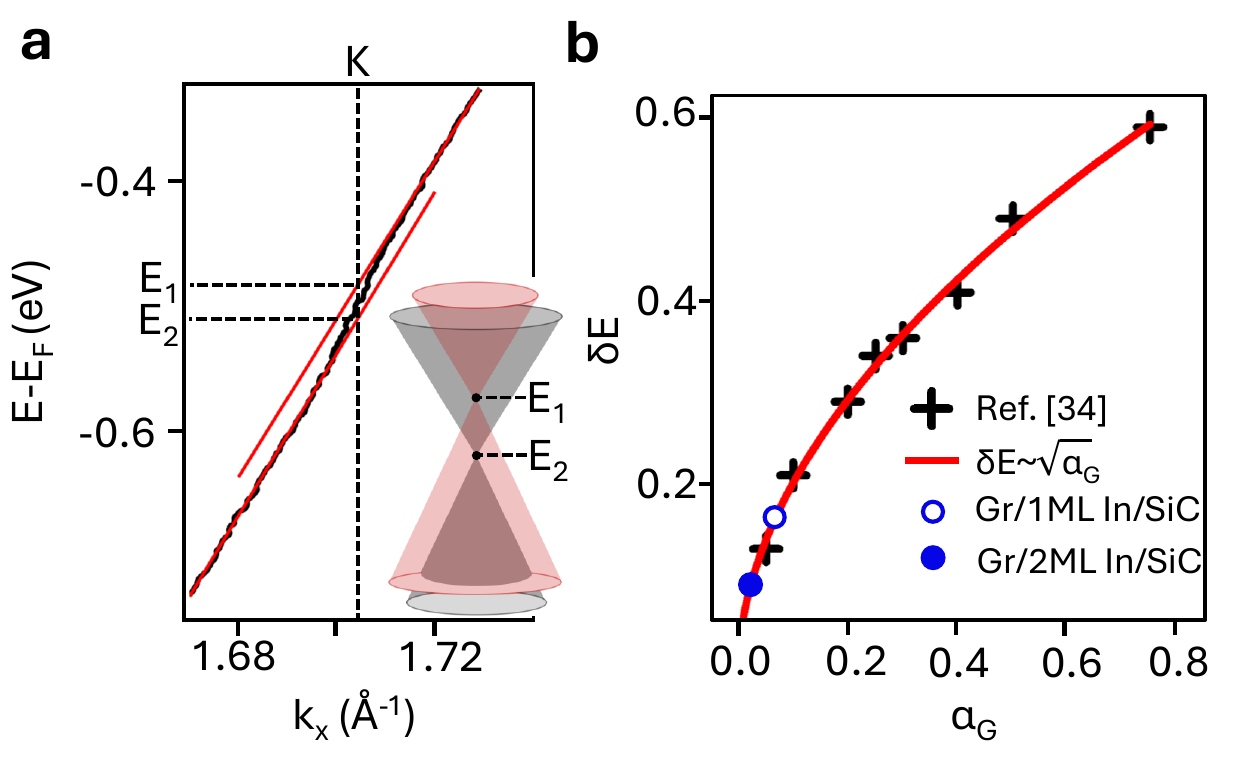}
\caption{\textbf{Graphene effective coupling constant.}
\textbf{a} Linear fits of Fig.~3\textbf{b\textsubscript{3}} above and below the plasmaron excitation used to extract the energy separation between the hole ($E_1$) and plasmaron ($E_2$) bands, as schematically illustrated in the inset. \textbf{c} Energy separation $\delta E$ as a function of the effective graphene coupling constant $\alpha_\mathrm{G}$, extracted from Ref.~\cite{Walter} (black cross). The red curve corresponds to an extrapolated fit to the data. The blue full (empty) circle highlights the energy separation in the bilayer (monolayer) indium-intercalated graphene.}
\label{fig:Graphene2}
\end{figure}

To quantify this behavior, momentum distribution curves (MDCs) within $-1~\mathrm{eV} < E-E_\mathrm{F} < 0~\mathrm{eV}$ were fitted with Lorentzians (white line in Fig.~\ref{fig:Graphene}b\textsubscript{3}). Three linear regions were identified: Region I ($0$ to $-0.1~\mathrm{eV}$), Region II ($-0.2$ to $-0.4~\mathrm{eV}$), and Region III ($-0.7$ to $-0.9~\mathrm{eV}$). Linear extrapolation of Regions III and II defines the crossing points ($E_1$ and $E_2$) of the hole and plasmaron bands, respectively. The normalized energy separation is $\delta E = (E_2 - E_1)/E_1$, with $E_1$ and $E_2$ denoting the energies of the fitted lines at $K_\mathrm{Gr}$, as depicted in Fig.~\ref{fig:Graphene2}a. Since the plasmon energy scales with the carrier density as $\omega_\mathrm{pl} \propto \sqrt{n}$ at fixed wave vector $q$ \cite{BOSTWICK200763}, this dependence can be accounted for by normalizing the energy difference with $E_1$, which exhibits the same $\sqrt{n}$ scaling \cite{ElectronDensityGrapheneFS}. As shown in Ref.~\cite{Walter}, the normalized energy separation is thus independent of the n-doping level. From this analysis, we obtain $\delta E = 0.0557 \pm 0.0022$, where the uncertainty was estimated using a min–max envelope analysis.

The effective graphene coupling constant $\alpha_\mathrm{G}$ is an indicator of the strength of Coulomb interactions between charge carriers. Following Ref.~\cite{Walter}, the relationship between $\delta E$ and $\alpha_\mathrm{G}$ is established through comparison with $G_0W_0$-RPA calculations~\cite{Polini}. The calculated $\delta E$ versus $\alpha_\mathrm{G}$ data are shown in Fig.~ \ref{fig:Graphene2}b. While the functional form of the underlying dependence is not known \textit{a priori}, an empirical fit ($\delta E (\alpha_\mathrm{G})=\alpha_\mathrm{G}^p$ with $p = 0.534 \pm 0.017$) provides an excellent description of the data ($R^2 = 0.995$) and allows for extrapolation of the graphene indium bilayer heterostructures (blue dots in Fig.~\ref{fig:Graphene2}b).

From this comparison, the effective coupling constant of graphene on bilayer indium is determined as $\alpha_\mathrm{G} = 0.0089 \pm 0.0007$. Using $\epsilon = e^2 / (4 \pi \epsilon_0 \alpha_\mathrm{G} \hbar v_\mathrm{F})$ and $\epsilon_s \approx 2\epsilon - 1$ \cite{Walter}, the intrinsic dielectric constant of graphene is $\epsilon = 312 \pm 25$ and the substrate screening is $\epsilon_s = 622 \pm 49$. These values provide direct insight into the dielectric environment: $\epsilon$ reflects the intrinsic response of the two-dimensional electron system, while $\epsilon_s$ captures the extrinsic contribution of the supporting substrate.

\begin{table}[h!]
\centering
\caption{Substrate effective screening constants $\epsilon_s$ of graphene Ref.~\cite{Walter,Au_Intercalation,Ag2,PhysRevB.105.235428}. $\delta E$ and $\delta k$ denote the energy and momentum separation of the hole and plasmaron bands, respectively.}
\label{tab:screening}
\begin{tabular}{lccc}
\hline\hline
Substrate & $\delta E$ & $\alpha_\mathrm{G}$ & $\epsilon_s$ \\
\hline
\textcolor{black}{Ag\textsubscript{(2)}-SiC} \cite{Ag2}      & $0.49 \pm 0.04$ & $0.5$ & $7.8$\\
H-SiC \cite{Walter}      & $0.49 \pm 0.02$ & $0.5 \pm 0.03$ & $7.8 \pm 0.5$\\
F-SiC \cite{Walter}      & $0.40 \pm 0.05$ & $0.4 \pm 0.05$& $10 \pm 1.3$ \\
\textcolor{black}{Ag\textsubscript{(1)}-SiC} \cite{PhysRevB.105.235428}      & $0.40 \pm 0.04$ & $0.4$ & $10.0 \pm 0.5$\\
\textcolor{black}{s}Au-SiC \cite{Au_Intercalation}   & $0.253$ & $0.3 \pm 0.02$ &$13 \pm 2$\\
ZLG-SiC \cite{Walter}    & $0.21 \pm 0.02$ &  $0.1 \pm 0.04$ &$43 \pm 16$\\
1ML In-SiC      & $0.17 \pm 0.01$ & $0.07 \pm 0.01$& $ 56.6 \pm 8.2$\\
\textcolor{black}{m}Au-SiC \cite{Walter}    & $0.12 \pm 0.04$ & $0.05 \pm 0.01$ & $87 \pm 18$\\
2ML In-SiC      & $0.056 \pm 0.002$ & $0.0089 \pm 0.0007$& $ 622 \pm 49$\\

\hline\hline
\end{tabular}
\end{table}

The bilayer indium–intercalated system exhibits an effective screening nearly an order of magnitude larger than that reported for previously studied systems, including graphene on monolayer indium (see Tab.~\ref{tab:screening}) \cite{Au_Intercalation,Walter}. Monolayer indium exhibits an effective screening of a similar magnitude to that observed in other intercalated materials. This highlights the importance of the second indium layer, which is necessary for the formation of nearly free-electron-like states leading to the high effective screening. This enhanced screening preserves the single-particle character of graphene and effectively suppresses the strong substrate interactions that have so far been unavoidable. 

\textcolor{black}{The comparison between the semiconducting sAu and metallic mAu-intercalated phases further underscores the decisive role of metallicity in determining the screening efficiency, with the metallic phase exhibiting substantially stronger dielectric screening \cite{Au_Intercalation,Walter}. Similar opportunities exist for Bi intercalation, where several experimentally realized phases span the range from insulating to metallic \cite{Sohn2021,Gehrig}. Likewise, Ag intercalation demonstrates that even within the two insulating phases, variations in atomic packing density can influence the screening properties \cite{Ag2,PhysRevB.105.235428}. Looking forward, Ga appears particularly promising, as bilayer and trilayer Ga are predicted to host similar metallic nearly free-electron states to those in bilayer indium, potentially providing equally strong dielectric screening. \cite{Briggs2020,Yi2026}.}


\noindent

In summary, we have demonstrated that graphene intercalated with bilayer indium exhibits pronounced many-body renormalizations in the Dirac state dispersion. In addition to the characteristic phonon-induced quasiparticle near the Fermi level, we resolve distinct electron–plasmon coupling that gives rise to plasmaron bands close to the Dirac point. From the energy separation between the hole and plasmaron bands, we extract the coupling constant $\alpha_\mathrm{G} = 0.0089 \pm 0.0007$ and quantify the resulting dielectric substrate screening $\epsilon_s = 622 \pm 49$, revealing an unusual strong screening environment provided by the bilayer indium. This behavior can be understood from the interplay of both indium layers: the first layer, exhibiting Rashba-like band splitting up to $161 \,$meV, buffers the SiC interaction, allowing the formation of nearly free-electron-like states in the second layer, which, in turn, enhances screening of the graphene layer.

While enhanced graphene screening marks a significant milestone for the graphene-on-SiC platform, its overall electronic performance remains limited by intrinsic substrate-related factors, such as SiC step height and density, and therefore still falls short of the regime achieved in suspended or fully encapsulated graphene \cite{ma16247668,BOLOTIN2008351,PhysRevLett.101.096802,Domaretskiy2025}, leaving direct transport measurements beyond the scope of the present ARPES study.
Our results highlight the controllability of graphene screening via the choice of intercalants, which provides an effective tuning knob for engineering its many-body interactions. Moreover, the scalability of this approach enables wafer-scale implementations, offering a promising route toward well-tailored graphene heterostructures for future high-speed electronic applications.

\section*{Methods}

\noindent N-doped 4H-SiC substrates were treated in a hydrogen dry-etching process yielding an atomically flat and oxygen-free ($1 \, \times \,1$) SiC(0001) surface \cite{Glass}. Subsequently, we grow the ($6\sqrt{3} \, \times \,6\sqrt{3}$)R30$^\circ$ reconstruction, also referred to as zero-layer graphene (ZLG), by heating SiC in argon atmosphere (950$\,$mbar) to 1365$^\circ$C for 10$\,$min \cite{Riedl_2010,ZLG_2,ZLG_3}. This ZLG serves as a substrate for the indium intercalation. 
Indium was intercalated by evaporating pure indium atoms ($\leq$99.9999$\%$) from a Knudsen cell. The intercalation itself is a two-step process \cite{Schmitt}. First, we evaporated the In atoms for 45$\,$min at room temperature (RT) and after that we annealed the sample to 500$^\circ$C for 20$\,$min. All the substrate temperature readings were done with an optical pyrometer from Keller with an emissivity of 85$\%$. For the LEED patterns the energy was set to 80$\,$eV.
The ARPES measurements were performed at the beamline I05 at Diamond Light Source (DLS). The measurements were taken with linear horizontal polarized light with a photon energy of 40$\,$eV at 10$\,$K, while the pressure was $\leq \, $3$\, \cdot \, 10^{-10}$mbar. 
\textcolor{black}{DFT calculations presented in Fig.~1c and 2b are performed within the
density functional theory framework as implemented in the
Vienna ab initio simulation package (VASP) within the projector-augmented plane- wave (PAW) method \cite{Kresse_1999,Blochl_1994}. For the exchange-correlation potential the PBE functional was used \cite{PBE}, by expanding the Kohn-Sham wave functions into plane-waves up to an energy cutoff of 400$\,$eV. 
We sampled the Brillouin zone on an 21$\times$21$\times$1 regular mesh with self-consistently included SOC \cite{Steiner_2016}.
We consider a bilayer indium on four layers of Si-terminated SiC(0001) with an in-plane lattice constant of 3.07$\,\text{\AA}$. The equilibrium structure was obtained by relaxing all atomic positions until the residual forces were below $0.005\,\mathrm{eV/\AA}$ \cite{Bauernfeind}. A vacuum distance of at least 20$\,\text{\AA}$ between periodic replicas in z-direction is applied.}


\bigskip
\textbf{Acknowledgements} 
We are grateful for funding support from the Deutsche Forschungsgemeinschaft (DFG, German Research Foundation) under Germany's Excellence Strategy through the W\"urzburg-Dresden Cluster of Excellence on Complexity, Topology and Dynamics in Quantum
Matter ctd.qmat (EXC 2147, Project ID 390858490) as well as through the
Collaborative Research Center SFB 1170 ToCoTronics (Project ID 258499086). We
acknowledge the Diamond Light Source for time at the beamline I05 (proposal SI40195).

\bigskip
\textbf{Data Availability} 
The data that support the findings of this article are openly available~\cite{WueData}.


{\noindent
	\textbf{Author contributions}
C.S. and L.G. have realized the sample growth and carried out the photoelectron spectroscopy experiments and their analysis. S.E. has conceived the theoretical ideas and performed the DFT calculations. On the experimental side, contributions came from K.S., J.E., M.K., T.K.,  J.S., S.M. and R.C. while G.S. gave inputs to the theoretical aspects. R.C. supervised this joint project and wrote the manuscript together with all other authors.
}

{\noindent
	\textbf{Competing interests}
	The authors declare no competing interests.
}

%

\end{document}